\documentclass[letterpaper,10pt,conference]{ieeeconf}
\IEEEoverridecommandlockouts
\usepackage{amsmath,amssymb,amsfonts}
\usepackage{cite}
\usepackage{graphicx}
\usepackage{subcaption}
\usepackage{textcomp}
\usepackage{xcolor}
\usepackage{comment}
\usepackage{mathtools}
\usepackage{tikz}
\definecolor{wakegreen}{rgb}{0,0.5,0}

\newcommand{\R}{\mathbb{R}}
\newcommand{\mcl}[1]{\mathcal{#1}}

\makeatletter
\def\@begintheorem#1#2{\trivlist
\item[\hskip\labelsep{\bfseries #1\ #2.}]\itshape}
\def\@opargbegintheorem#1#2#3{\trivlist
\item[\hskip\labelsep{\bfseries #1\ #2\ (#3).}]\itshape}
\makeatother
\newtheorem{theorem}{Theorem}

\newtheorem{lemma}[theorem]{Lemma}
\newtheorem{definition}{Definition}

\title{\LARGE \bf 
State-Dependent Delay Modeling of Wind-Farm Wake Dynamics
}
\author{Sengiyumva Kisole$^{1}$, Jungbae Chun$^{2}$, Peter Seiler$^{2}$, and Matthew M.~Peet$^{1}$%
\thanks{This material is based upon work supported by the National Science Foundation under Grants NSF EPCN-2337751 and NSF EPCN-2337752.}%
\thanks{$^{1}$School for the Engineering of Matter, Transport and Energy, Arizona State University, Tempe, AZ 85298, USA.
{\tt\small sengi.kisole@asu.edu, mpeet@asu.edu}}%
\thanks{$^{2}$Department of Electrical Engineering and Computer Science, University of Michigan, Ann Arbor, MI 48109, USA.
{\tt\small jungbaec@umich.edu, pseiler@umich.edu}}%
}

\begin{document}
\maketitle
\thispagestyle{empty}
\pagestyle{plain}

\begin{abstract}

Turbines convert wind to electrical power. In the wake of a turbine, wind speed is reduced in proportion to the amount of power extracted. In a wind farm, downstream turbines may be located in wake of an upstream turbine, requiring coordination to yield maximum power generation. This paper creates a state-dependent delay model for a two turbine wind farm. The upstream turbine creates a wake that diminishes the wind speed at the downstream turbine. Changing the upstream turbine operation will affect the downstream turbine after a delay that depends on the wind-propagation time between them. This delay depends on the upstream power extraction and it can destabilize the combined dynamics. Conventional models of this effect rely on nonlinear partial differential equations (PDEs). Control design and analysis with such models is difficult with established methods. We first show that the PDE model can be represented by an Abel-type state-dependent delay differential equation. The equation has an implicit solution. To obtain an explicit representation, we then introduce a simplified state-dependent delay model. Both models predict destabilizing interactions between the two turbines. Numerical examples illustrate delay-induced oscillations using the simplified model and verify their effectiveness in simulation.

\end{abstract}
\section{Introduction} \label{sec:introduction}

A wind turbine converts a portion of the kinetic energy in the wind into mechanical energy at the rotor, and its generator then converts that mechanical energy into electrical energy. 
%
%
However, locations with strong and persistent wind resources are limited and require access to an electrical distribution network, access roads, and maintenance resources.  Furthermore, limitations include: lease boundaries, terrain, environmental impact, and permitting requirements. As a result, wind turbines are often aggregated into wind farms in which upstream turbines may reduce the available wind energy available at downstream turbines~\cite{boersma2017tutorial,pedersen2020integrated}.

To minimize interaction between turbines, wind-farm layout-optimization may be used to position turbines by accounting for geography and measured wind-direction profiles~\cite{goccmen2016wind,pedersen2020integrated}. However, even in optimal wind-farm layouts, wake losses due to turbine wake interaction are inevitable~\cite{johnson2009wind,boersma2017tutorial}. To maximize power generation for a given wind farm layout, coordinated control can be used. While optimal operation of an individual turbine captures  maximum power, when wake interactions are present such a strategy is suboptimal\cite{bitar2013coordinated, johnson2009wind}. Coordinated control, then, throttles upstream turbines in order to increase generation at downstream turbines by aggregating wind data from all turbines and designing reference operating points. Feedback controllers then stabilize individual turbines about their prescribed operating points~\cite{bitar2013coordinated}. However, the resulting network feedback control architecture must be designed carefully to ensure stability. This stabilization problem is particularly challenging due to delays induced by wake propagation from upstream to downstream turbines. 

To illustrate the challenge of ensuring stability of a coordinated wind-farm, consider a simple two-turbine model with Turbine~1 upstream of Turbine~2.  
Due to wake interactions, an increase in rotor speed at Turbine 1 will result in a decreased wind speed and hence decreased downstream power generation. In steady-state coordinated control, desired rotor speeds, $\omega_1,\omega_2$ are chosen to maximize total power generation. The challenge, however, is stabilization about this desired set-point in the presence of time-varying aerodynamic torque. This problem is complicated by the fact that aerodynamic torque at the downstream Turbine 2 will vary with changes in the rotor speed of the upstream turbine. Meanwhile, feedback control at Turbine 1 may attempt to compensate for deviations from optimality at downstream turbine by reducing its generator torque. This scenario is conceptually similar to the stabilization problems faced in feed-forward control of vehicle platoons\cite{al2017feedforward,seiler2004disturbance}. The primary difficulty for wind turbines, however, is modeling how changes in rotor speed at the upstream turbine propagate to changes in aerodynamic torque at the downstream turbine. Specifically, this effect is not instantaneous and is governed by PDEs which represent the spatially distributed process of wake expansion and propagation. 


Existing wake models differ in their ability to capture how changes in the rotor speed at an upstream turbine propagate to changes in the aerodynamic torque at a downstream turbine. Steady state wake models such as Jensen-Park model~\cite{katic1987simple} by definition neglect time-variation. By contrast, nonlinear PDE wake models such as in~\cite{shapiro2019wake,shapiro2017model} explicitly model the evolution of the distributed wind velocity as it propagates between turbines. The difficulty in using such models to design stabilizing feedback controllers lies in the limited availability of tools for analyzing the stability of nonlinear PDEs. Furthermore, from the perspective of control design, by modeling the wake at all points between turbines, PDE wake models provide more information than is necessary to describe the effect of changes in wake at a discrete point (upstream turbine) on changes in wind velocity at another discrete point (downstream turbine). The primary goal of this paper, then, is to ignore evolution of the intermediate evolution of the wake and obtain a more direct relationship between upstream and downstream wind velocity. 

The problem of reducing transport PDEs to simplified discrete input-output relations has been studied in a variety of contexts, including maturation in biological systems~\cite{smith1993reduction} and manufacturing~\cite{diagne2017time}. Specifically, it has been shown that many simple PDE models including linear transport equations can be reduced to discrete-delay models using the method of characteristics~\cite{karafyllis2014relation,bekiaris2018compensation}. For the nonlinear PDE transport equations used to model wake propagation, similar approaches can be taken~\cite{diagne2017control}. However, in this case, the delay is time-varying and implicitly defined. In this paper, we apply these methods to nonlinear wake propagation and similarly obtain a time-varying delay model where the delay is implicitly defined using the solution to a functional Abel type equation -- relating wake upstream at time $t_0$ ( i.e. $v(t_0)$) to wake downstream at time $t_0+\tau(v(t_0,0))$. Numerical simulations are used to verify that this model accurately captures evolution of the nonlinear PDE as discrete points in the wake. 

The difficulty with implicitly defined delay models, however, is that they require us to obtain fixed-points of an Abel-type functional equation. Control of such systems then requires to the use of robust control, integral relationships, or back-stepping~\cite{diagne2017control}. The approach taken in this paper is somewhat different, however. Specifically, we use the implicit inversion of the Abel-type functional equation to define a time-varying delay model. We then integrated this into the turbine dynamics to obtain a state-dependent delay model where the state-dependent delay can be constructed using an approximation to the inverse of the Abel-type functional equation. The model is then tested numerically and shown to accurately capture instabilities arising from aerodynamic coupling between turbines.

The paper is organized as follows. The wake models—the Jensen--Park model, the transport PDE model, Abel-type functional model, and the state-dependent delay models—are introduced in Sec.~\ref{Sec:WakeModeling}. The two-turbine rotor dynamics and generator-torque feedback laws about the operating points are given in Sec.~\ref{Sec:RotorDynamicsandControl}. The coupling between the wake models and rotor dynamics is formulated in Sec.~\ref{Sec:CoupledSystem}. Finally, the rotor-speed and downstream-inflow responses are compared in Sec.~\ref{Sec:NumericalResults} for two initial conditions, with operating points and feedback gains held fixed to isolate the effect of wake propagation and assess agreement with the PDE response.

\section{Problem Formulation}
\subsection{On the Wake Model}
Consider a simple model of wind flow along a single direction. Let $v(t,x)$ denote the wind speed at time $t$ and location $x$. The goal is to model the effect of an upstream velocity $v(t,0)$ on the the downstream velocity $v(t,L)$ where $L>0$. 
Our starting point is the PDE model of the form:
\vspace{-2mm}
\begin{equation}\label{Eqn:ProblemPDE}
\frac{\partial v}{\partial t}
+v(t,x)\frac{\partial v}{\partial x}
=2v(t,x)\frac{D_w'(x)}{D_w(x)}
\bigl[v_1-v(t,x)\bigr].
\tag{M1}
\end{equation}
The boundary velocity is given by a function $v_0:\R\to\R$, i.e., $v(t,0)=v_0(t)$.  Moreover, $v_1$ is the constant free-stream velocity and $D_w(x):= D+2 k_w x$ describes the dimension of the wake at location $x$.  Here, the rotor diameter $D$, and wake expansion factor $k_w$, are model constants.

We can use the method of characteristics to obtain a solution to \eqref{Eqn:ProblemPDE} of the form:
\begin{equation}\label{Eqn:ProblemForwardPropagation}
v\bigl(t + \tau(v(t,0)),L\bigr)
=\beta(L)v(t,0)+[1-\beta(L)]v_1.
\tag{M2}
\end{equation}
Here, $\beta(x):=\left(D/D_w(x)\right)^2$, and
\[
\tau(z):=\int_0^L
\frac{dx}{v_1-(v_1-z)\beta(x)}, \qquad 0<z\leq v_1.
\]

Let $\theta(t):=t +\tau(v(t,0))$. We assume the flow satisfies a first-in, first-out behavior: $\dot{\theta}(t)>0$, $\theta$ is strictly increasing and hence invertible. We then invert $\theta$ to obtain an implicit state-dependent delay model:
\begin{equation}\label{Eqn:ProblemStateDependentWake}
v(t,L)= \beta(L)v\bigl(t-\tau\bigl(v(\theta^{-1}(t),0)\bigr),0\bigr)
+[1-\beta(L)]v_1,
\tag{M3}
\end{equation}
for all $t\geq\theta(0)$.

We then assume that the delay is  slow varying, $\left|\tau'(v_2(t,0))\,\partial_t v_2(t,0)\right|\ll1$, and  approximate $\tau(v_2(\theta^{-1}(t),0))$ by $\tau(v_2(t,0))$, to obtain an explicit state-dependent delay model
\begin{equation}\label{Eqn:ExplicitWakeProb}
v_2(t,L)=\beta(L)v_2\bigl(t-\tau(v_2(t,0)),0\bigr)+[1-\beta(L)]v_1.
\tag{M4}
\end{equation}
\subsection{On the Rotor Dynamics}
Having obtained our wake model, we integrate it into the rotor dynamics of a  two-turbine wind farm operating about an reference set point to obtain the state-dependent delay differential equation:
\begin{equation}
\begin{aligned}
&I_1\dot{\omega}_1(t)=\frac{\rho A v_1^3}{2\omega_1(t)}
C_P\!\left(\frac{R\omega_1(t)}{v_1},\hat\beta\right)-T_{g,1}(t),\\
&I_2\dot{\omega}_2(t) =\frac{\rho A h_2^3(t)}{2\omega_2(t)}
C_P\!\left(\frac{R\omega_2(t)}{h_2(t)},\hat\beta\right)-T_{g,2}(t),\\
&h_2(t)=\beta(L)g\bigl(t-\tau(g(t))\bigr)+[1-\beta(L)]v_1,
\end{aligned}
\tag{M5}
\end{equation}
with
\vspace{-2mm}
\[
g(t) = v_1\sqrt{1-C_T\left(\frac{R\omega_1(t)}{v_1},\hat\beta\right)}, 
\]
and coordinated control laws:
\begin{align*}
T_{g,1}(t)&=T_{g,1}^*+K_1\bigl(\omega_1(t)-\omega_1^*\bigr)
+K_2\bigl(\omega_2(t)-\omega_2^*\bigr), \\
T_{g,2}(t)&=T_{g,2}^*+K_3\bigl(\omega_2(t)-\omega_2^*\bigr),
\end{align*}
where $\omega_1^*,\omega_2^*$ are reference operating points, and $T_{g,1}^*$ and $T_{g,2}^*$ are generator-torques at the reference operating points ,  and $K_1,K_2,K_3$ are feedback gains. 

Given the time-varying upstream boundary velocity $v_2(t,0)$, which depends on $\omega_1(t)$, our goal is to determine the downstream inflow $v_2(t,L)$ while accounting for the finite propagation of wake effects over the turbine spacing $L$. We aim to derive the state-dependent delay $\tau(v_2(t,0))$ from a dynamic PDE wake model so that it can be combined with the rotor dynamics for stability analysis and control design.

\section{Wake Modeling}\label{Sec:WakeModeling}
\subsection{Steady-State Jensen--Park Wake Model}\label{Sec:JensenWakeModel}

In this Sec. we first introduce the Jensen--Park model describes the spatial variation of the wake-velocity deficit generated by a turbine operating at a fixed tip-speed ratio~\cite{katic1987simple,goccmen2016wind}, while the transport PDE describes how time-varying changes at $x=0$ propagate  downstream to $x=L$. The model assumes a uniform velocity across each wake cross section, linear wake expansion in the downstream direction, and that Turbine~1 operate at the fixed rotor speed $\omega_1$ and constant free-stream inflow $v_1$. 

Let $v_w(x;\omega_1)$ denote the wake velocity at a distance $x$ downstream of an upstream turbine, Turbine~1. 
\begin{definition}\label{JensenWakeDefn}
Given turbine spacing $L>0$, free-stream velocity $v_1>0$ and a fixed upstream rotor speed $\omega_1$,
$\Delta v(\cdot;\omega_1):[0,L]\rightarrow[0,v_1)$ is said to be a \textbf{steady-state wake-velocity deficit} if  for all $x\in[0,L]$
\vspace{-2mm}
\begin{equation}\label{Eqn:JensenWakeDeficitDefn}
\Delta v(x;\omega_1)
=v_1-v_w(x;\omega_1),
\end{equation}
where $v_w(\cdot;\omega_1)$ is the wake-velocity.
\end{definition}

The axial-induction factor $a_1(\omega_1)$ measures the fractional decrease in axial wind speed from the free stream to Turbine~1's rotor plane.
For a fixed blade-pitch angle $\hat\beta$, actuator-disk theory gives
\vspace{-2mm}
\begin{equation}\label{Eqn:AxialInductionFactor}
4a_1(\omega_1)\bigl(1-a_1(\omega_1)\bigr)=
C_T\left(R\omega_1/ v_1,\hat\beta\right),
\end{equation}
where $C_T$ is the thrust coefficient.
The wake diameter expands as
\(
D_w(x)=D+2k_wx,
\)
where $D=2R$ and $k_w>0$ is the wake-expansion coefficient. The deficit decreases with the square of the wake-diameter expansion:
\vspace{-2mm}
\begin{equation}\label{Eqn:JensenWakeDeficit}
\frac{\Delta v(x;\omega_1)}{v_1}
=2a_1(\omega_1)\left(\frac{D}{D_w(x)}\right)^2.
\end{equation}
Hence, by  Defn.~\ref{JensenWakeDefn}, the wake velocity is
\vspace{-2mm}
\begin{equation}\label{Eqn:JensenWakeVelocity}
v_w(x;\omega_1)
=v_1\left[1-2a_1(\omega_1)
\left(\frac{D}{D_w(x)}\right)^2\right].
\end{equation}

The Jensen--Park wake model determines the downstream inflow associated with a fixed upstream operating state, however, it does not describe how wakes propagate downstream. To model how time-varying upstream operation affects downstream inflow, let $v_2(t,x)$ denote the velocity between turbines. An advection equation describes the downstream motion of the flow at the local wake speed as the wake expands; the upstream rotor speed determines the velocity at the boundary $x=0$. The resulting inflow velocity at a downstream turbine $v_2(t,L)$ accounts for the time it takes for the wake to propagate providing a dynamic
transport wake model.

\subsection{From Wake Transport to a  Abel-type Functional Equation}\label{Sec:AbelWakeModel}
Consider the wake generated by Turbine~1 over the interval $x\in[0,L]$, where Turbine~1 is located at $x=0$ and Turbine~2
is located at $x=L$. Let $v_2(t,x)>0$ denote the local wake velocity at time $t$ and position $x$. The free-stream velocity
$v_1>0$ is assumed to be constant.

Suppose the wake expands downstream with diameter
\vspace{-2mm}
\[
D_w(x)=D+2k_wx,
\]
where $D$ is the rotor diameter and $k_w>0$ is the wake-expansion coefficient.
Then the cross-sectional area of the wake is proportional to $D_w^2(x)$, and the wake expansion causes the local velocity
to recover toward the free-stream velocity as the wake travels downstream.

\textbf{Model 1:}  Assume that the wake propagates downstream with convection velocity $v_2(t,x)$. 
As the wake diameter increases, $v_2(t,x)$ recovers toward the free-stream velocity $v_1$. 
Then, the local wake velocity satisfies the transport equation
\vspace{-2mm}
\begin{equation}\label{Eqn:DynamicWakeVelocityPDE}
\frac{\partial v_2}{\partial t}
+v_2(t,x)\frac{\partial v_2}{\partial x}
=2v_2(t,x)\frac{D_w'(x)}{D_w(x)}
\left[v_1-v_2(t,x)\right],
\end{equation}
where $\partial v_2/\partial t$ is the local rate of change, $v_2(t,x)\partial v_2/\partial x$ is the convection,
and $2v_2(t,x)D_w'(x)\left[v_1-v_2(t,x)\right]/D_w(x)$ is the wake-recovery term w.r.t. the wake diameter.

Since 
\vspace{-2mm}
\[
2\frac{D_w'(x)}{D_w(x)}=\frac{4k_w}{D+2k_wx},
\]
the wake recovery term is nonnegative whenever $0<v_2(t,x)\leq v_1$.

The wake generated by an upstream turbine takes time to propagate to a downstream turbine, delaying its effect on the downstream inflow. Our goal is to determine this transport delay. To do this, we solve the transport equation using the method of characteristics, which reduces the PDE to an ordinary differential equation along each downstream path. Solving the ordinary differential equation with the upstream boundary condition gives the wake velocity along each path. Integrating the reciprocal of the wake velocity over the distance between the turbine gives the wake it takes to travel downstream. 

Lemma~\ref{Lem.3} derives a forward-time Abel-type functional equation from the wake transport PDE~\eqref{Eqn:DynamicWakeVelocityPDE}, relating a upstream boundary velocity at time $t_0$ to a downstream inflow at time $t_0+\tau(v_2(t_0,0))$, but first we define a characteristic curve of a PDE.

\begin{definition}\label{Def:CharacteristicCurve}
Given functions $f,g,h: \R^3 \rightarrow \R$.  A continuously differentiable curve  $s \mapsto (x(s),t(s),u(x(s),t(s)))$ is said to be a \textbf{characteristic curve} associated with $f,g,h$  if for all $s$,
\vspace{-3mm}
\[
\begin{aligned}
\frac{d}{ds}x(s)&=g(x(s),t(s),u(x(s),t(s))),\\
\frac{d}{ds}t(s)&=h(x(s),t(s),u(x(s),t(s))),\\
\frac{d}{ds}u(x(s),t(s))&=f(x(s),t(s),u(x(s),t(s))),
\end{aligned}
\]
with $x(0)=0$, $t(0)=t_0$, and
$u(x(0),t(0))=\phi(t_0)$.
\end{definition}

\begin{lemma}\label{Lem.3}
Given $L,D,k_w,v_1>0$, for all $x\in[0,L]$, define
\[
D_w(x):=D+2k_wx.
\]
Suppose $v_2$ is continuously differentiable on
$[0,\infty)\times[0,L]$ and satisfies
\begin{equation}\label{Eqn:PDE02}
\frac{\partial v_2}{\partial t}
+v_2(t,x)\frac{\partial v_2}{\partial x}
=2v_2(t,x)\frac{D_w'(x)}{D_w(x)}
[v_1-v_2(t,x)],
\end{equation}
with $0<v_2(t,0)\leq v_1$ for all $t\geq0$. Define
\vspace{-2mm}
\begin{equation}\label{Defn:Beta}
\beta(x):=\left(\frac{D}{D_w(x)}\right)^2,
\qquad x\in[0,L],
\end{equation}
and
\vspace{-2mm}
\begin{equation}\label{Defn:WakePropagationTime}
\tau(z):=\int_0^L
\frac{dx}{v_1-(v_1-z)\beta(x)},
\qquad 0<z\leq v_1.
\end{equation}
Then, for all $t_0\geq0$,
\begin{equation}\label{Eqn:ForwardWakePropagation}
v_2\bigl(t_0+\tau(v_2(t_0,0)),L\bigr)
=\beta(L)v_2(t_0,0)+[1-\beta(L)]v_1.
\end{equation}
\end{lemma}

\begin{proof}
We first show that $[v_1-v_2(t(s),x(s))]D_w^2(x(s))$ is constant along the characteristic curve and evaluate it at $s=0$.
We then integrate the reciprocal velocity to obtain the time it takes for the wake to propagate downstream.

For any $t_0\geq0$, let
\vspace{-2mm}
\[
\begin{aligned}
f(x,t,z)&:=2z\frac{D_w'(x)}{D_w(x)}(v_1-z),\\
g(x,t,z)&:=z,\\
h(x,t,z)&:=1.
\end{aligned}
\]
With $u(x,t)=v_2(t,x)$ and $\phi(t)=v_2(t,0)$, Eqn.~\eqref{Eqn:PDE02} has the form $gu_x+hu_t=f$.

By Defn.~\ref{Def:CharacteristicCurve}, a characteristic curve $s\mapsto\bigl(x(s),t(s),v_2(t(s),x(s))\bigr)$ associated with $f,g,h$ satisfies
\vspace{-2mm}
\begin{equation}\label{Eqn:WakePath}
\begin{aligned}
\frac{d}{ds}x(s)&=v_2(t(s),x(s)),\\
\frac{d}{ds}t(s)&=1,
\end{aligned}
\end{equation}
and
\begin{equation}\label{Eqn:WakeVelocityPathODE}
\begin{aligned}
&\frac{d}{ds}v_2(t(s),x(s))\\
&\quad=\frac{2D_w'(x(s))}{D_w(x(s))}
v_2(t(s),x(s))[v_1-v_2(t(s),x(s))],
\end{aligned}
\end{equation}
with $x(0)=0$, $t(0)=t_0$, and initial velocity $v_2(t_0,0)$.
Integrating the second equation in \eqref{Eqn:WakePath} gives $t(s)=t_0+s$.

Since $v_1$ is constant, the product rule gives
\begin{equation}\label{Eqn:WakePathProductRule}
\begin{aligned}
&\frac{d}{ds}\left\{
[v_1-v_2(t(s),x(s))]D_w^2(x(s))
\right\}\\
&\quad=-D_w^2(x(s))\frac{d}{ds}v_2(t(s),x(s))\\
&\qquad+2[v_1-v_2(t(s),x(s))]
D_w(x(s))D_w'(x(s))\frac{d}{ds}x(s).
\end{aligned}
\end{equation}

Using \eqref{Eqn:WakeVelocityPathODE}, the first term on the right-hand side of \eqref{Eqn:WakePathProductRule} simplifies to
\vspace{-2mm}
\[
\begin{aligned}
&-D_w^2(x(s))\frac{d}{ds}v_2(t(s),x(s))\\
&\quad=-2D_w(x(s))D_w'(x(s))
v_2(t(s),x(s))[v_1-v_2(t(s),x(s))].
\end{aligned}
\]
Using \eqref{Eqn:WakePath}, the second term on the right-hand side of \eqref{Eqn:WakePathProductRule} simplifies to
\vspace{-3mm}
\[
\begin{aligned}
&2[v_1-v_2(t(s),x(s))]
D_w(x(s))D_w'(x(s))\frac{d}{ds}x(s)\\
&\quad=2D_w(x(s))D_w'(x(s))
v_2(t(s),x(s))[v_1-v_2(t(s),x(s))].
\end{aligned}
\]
Substituting both expressions into \eqref{Eqn:WakePathProductRule} gives
\vspace{-2mm}
\begin{equation}\label{Eqn:WakePathInvariant}
\frac{d}{ds}\left\{
[v_1-v_2(t(s),x(s))]D_w^2(x(s))
\right\}=0.
\end{equation}

Integrating \eqref{Eqn:WakePathInvariant} from $0$ to $s$ and using $x(0)=0$, $t(0)=t_0$, and $D_w(0)=D$, we obtain
\begin{equation}\label{Eqn:WakePathBoundaryIdentity}
[v_1-v_2(t(s),x(s))]D_w^2(x(s))=[v_1-v_2(t_0,0)]D^2.
\end{equation}

Since $D_w(x(s))\geq D>0$, dividing \eqref{Eqn:WakePathBoundaryIdentity} by $D_w^2(x(s))$
and using Eqn.~\eqref{Defn:Beta} gives
\begin{equation}\label{Eqn:WakeVelocityAlongPath}
v_2(t(s),x(s))
=\beta(x(s))v_2(t_0,0)+[1-\beta(x(s))]v_1.
\end{equation}

By \eqref{Defn:Beta}, $0<\beta(x(s))\leq1$ for $x(s)\in[0,L]$.
Since $0<v_2(t_0,0)\leq v_1$, Eqns.~\eqref{Eqn:WakePath} and \eqref{Eqn:WakeVelocityAlongPath} give
\(
v_2(t_0,0)\leq\frac{d}{ds}x(s)\leq v_1.
\)
Thus, $x(s)$ is strictly increasing and reaches $L$ in finite time.

Using \eqref{Eqn:WakePath} and \eqref{Eqn:WakeVelocityAlongPath}, integrating the reciprocal velocity along the characteristic gives
\[
s=\int_0^{x(s)}
\frac{dx}{v_1-[v_1-v_2(t_0,0)]\beta(x)}.
\]

At $x(s)=L$, Eqn.~\eqref{Defn:WakePropagationTime} gives $s=\tau(v_2(t_0,0))$.
Since $t(s)=t_0+s$, evaluating \eqref{Eqn:WakeVelocityAlongPath} at $x(s)=L$ gives
\[
v_2\bigl(t_0+\tau(v_2(t_0,0)),L\bigr)
=\beta(L)v_2(t_0,0)+[1-\beta(L)]v_1,
\]
as desired.
\end{proof}

Lem.~\ref{Lem.3} gives an Abel-type functional equation wake model in which the upstream rotor speed determines the time argument of the downstream velocity, making it difficult to obtain $v_2(t,L)$ explicitly. Abel-type functional equations associated with time transformations also involve function composition, under suitable conditions,  solutions  to Abel-type functional equations can be constructed by iterative propagation of seed functions~\cite{11615921,11616383}.

This provides our second model relating $v_2(t,L)$ and $v_2(t,0)$

\textbf{Model 2:} Given $v_1,D,D_w$, let $\tau(x)$ and $\beta (x)$ be as defined in Lemma~\ref{Lem.3}. Then 
\[
\beta(L)v_2(t_0,0)+[1-\beta(L)]v_1=v_2\bigl(t_0+\tau(v_2(t_0,0)),L\bigr)
\]
This model is transcendental and is similar in form to Abel's iterative functional equation -- containing a forward-in-time shift. Solutions to such equations do not admit direct representation using backward delays -- making them difficult to analyze using existing Lyapunov-based analysis. Consequently, in the following section, we use an implicit solution of the Abel functional equation~\eqref{Eqn:ForwardWakePropagation} to develop a backward-delay model which approximates solutions of Model 2.

\subsection{State-Dependent Delay Model} \label{Sec:StateDependentDelayModel}
Lem.~\ref{Lem.3} relates the wake velocity at the upstream boundary at time $t_0$ to the downstream inflow at time $t_0+\tau(v_2(t_0,0))$. To examine how wake propagation affects the coupled rotor dynamics, we express the downstream inflow at the current time in terms of the earlier boundary velocity. 
Suppose that
\(
1+\tau'(v_2(t_0,0)) \partial_t v_2 (t_0,0)>0
\)
for all $t_0\geq0$. Then the lag, $\theta(t_0)=t_0+\tau(v_2(t_0,0))$, is strictly increasing: characteristics entering at $x=0$ earlier reach $x=L$ earlier. The condition imposes first-in, first-out (FIFO) ordering. Inverting $\theta$ gives an implicit delay wake model:

\begin{theorem}
Given $L,D,k_w,v_1>0$, let $\tau\in\mcl C^1((0,v_1],\R_+)$ be defined by \eqref{Defn:WakePropagationTime}. Let $D_w(x):=D+2k_wx$ and let $\beta$ be defined by \eqref{Defn:Beta}. Suppose $v_2$ is continuously differentiable on $[0,\infty)\times[0,L]$ and satisfies \eqref{Eqn:PDE02}   and for all $t_0\geq0$, 
\(
1+\tau'(v_2(t_0,0)) \partial_t v_2(t_0,0)>0
\)
with $0<v_2(t_0,0)\leq v_1$
Define $\theta(t_0):=t_0+\tau(v_2(t_0,0))$. 

Then  for all $t\geq\theta(0)$, 
\vspace{-2mm}
\begin{equation}\label{Eqn:ExactDelayedDownstreamInflow}
v_2(t,L)=\beta(L)v_2\bigl(\theta^{-1}(t),0\bigr)+[1-\beta(L)]v_1.
\end{equation}
\end{theorem}

\begin{proof}
First, we show that $\theta$ is strictly increasing and maps $[0,\infty)$ onto $[\theta(0),\infty)$. For each $t\geq\theta(0)$, we then use $\theta^{-1}(t)$ to find the earlier time $t_0$ and apply Lem.~\ref{Lem.3} at $t_0$.

Since $\tau\in\mcl C^1((0,v_1],\R_+)$ and $v_2$ is continuously differentiable, $\theta$ is differentiable. Since
\vspace{-2mm}
\[
\dot{\theta}(t_0)=1+\tau'(v_2(t_0,0))\frac{\partial v_2}{\partial t}(t_0,0)>0
\]
for all $t_0\geq0$, we have that $\theta$ is strictly increasing on $[0,\infty)$.

Since $\tau(v_2(t_0,0))>0$, we have $\theta(t_0)>t_0$. Hence $\theta(t_0)\to\infty$ as $t_0\to\infty$. Thus, the range of $\theta$ is $[\theta(0),\infty)$, so $\theta^{-1}(t)$ is well-defined for every $t\geq\theta(0)$.

Let $t\geq\theta(0)$ and define $t_0=\theta^{-1}(t)$. Then
\vspace{-2mm}
\[
t=\theta(t_0)=t_0+\tau(v_2(t_0,0)).
\]

Finally, applying \eqref{Eqn:ForwardWakePropagation} from Lem.~\ref{Lem.3} at $t_0=\theta^{-1}(t)$, we obtain
\vspace{-2mm}
\[
v_2(\theta(t_0),L)=\beta(L)v_2(t_0,0)+[1-\beta(L)]v_1.
\]
Since $\theta(t_0)=t$ and $t_0=\theta^{-1}(t)$, we have
\[
v_2(t,L)=\beta(L)v_2\bigl(\theta^{-1}(t),0\bigr)+[1-\beta(L)]v_1.
\]
\end{proof}

\textbf{Model 3:} Given $L,D,k_w,v_1>0$, let $\tau\in\mcl C^1((0,v_1],\R_+)$ be defined by \eqref{Defn:WakePropagationTime}. Let $D_w(x):=D+2k_wx$ and let $\beta$ be defined by \eqref{Defn:Beta}. Under the FIFO condition $1+\tau'(v_2(t,0))\,\partial_t v_2(t,0)>0$, we have  
\begin{equation}
v_2(t,L)=\beta(L)v_2\bigl(\theta^{-1}(t),0\bigr)+[1-\beta(L)]v_1.
\end{equation}

\textbf{Model 4:}
From $\theta(t)=t+\tau(v_2(t,0))$, we have $\theta^{-1}(t)=t-\tau(v_2(\theta^{-1}(t),0))$.
Suppose there exists $0<\epsilon<1$, with $\epsilon\ll1$, such that
\[
\left|\tau'(v_2(r,0))\,\partial_t v_2(r,0)\right| \leq\epsilon
\]
for all $r\geq0$.
By the Mean Value Thm., for each $t\geq\theta(0)$, there exists
$\xi\in(\theta^{-1}(t),t)$ such that
\begin{align*}
&\left|\tau(v_2(t,0)) -\tau(v_2(\theta^{-1}(t),0))\right|\\
&\qquad \qquad = \left|\tau'(v_2(\xi,0))\,\partial_t v_2(\xi,0)\right|
\bigl(t-\theta^{-1}(t)\bigr)\\
&\qquad \qquad \qquad \leq\epsilon\bigl(t-\theta^{-1}(t)\bigr)=\epsilon\tau(v_2(\theta^{-1}(t),0)).
\end{align*}
Since $\epsilon\ll1$, we approximate $\tau(v_2(\theta^{-1}(t),0))$ by $\tau(v_2(t,0))$, and hence
$\theta^{-1}(t)$ by $t-\tau(v_2(t,0))$, to obtain an explicit
state-dependent delay model
\begin{equation}\label{Eqn:ExplicitSDModel}
v_2(t,L)=\beta(L)v_2\bigl(t-\tau(v_2(t,0)),0\bigr) +[1-\beta(L)]v_1.
\end{equation}

To determine the aerodynamic coupling due to wake interactions we substitute the downstream inflow velocity at the downstream rotor $v_2(t)$ into the turbine torque balance equation. In the next, section we consider a simple two-turbine model with Turbine~1 upstream of Turbine~2. 


\section{Two-Turbine Configuration and Aerodynamic Coupling}\label{Sec:RotorDynamicsandControl}

Consider a variable-speed wind turbine operating at the fixed blade-pitch angle $\hat\beta$. The rotor-speed dynamics are described by the torque balance~\cite{johnson2004stability,johnson2006control}
\vspace{-2mm}
\begin{equation}\label{Eqn:SingleTurbineDynamics}
I\dot{\omega}(t)=T_a(t)-T_g(t),
\end{equation}
where $I>0$ is the rotor inertia, $\omega(t)>0$ is the rotor speed, $T_a(t)$ is the aerodynamic torque, and $T_g(t)$ is the generator torque, which serves as the control input. For an inflow velocity $v(t)>0$, the aerodynamic torque is
\begin{equation}\label{Eqn:SingleTurbineAerodynamicTorque}
T_a(t)=\frac{\rho A v^3(t)}{2\omega(t)}
C_P\!\left(\frac{R\omega(t)}{v(t)},\hat\beta\right),
\end{equation}
where $\rho>0$ is the air density, $R>0$ is the rotor radius, $A=\pi R^2$ is the swept area, and
\vspace{-2mm}
\[
\lambda(t):=\frac{R\omega(t)}{v(t)}
\]
is the tip-speed ratio. The functions $C_P(\lambda,\hat\beta)$ and $C_T(\lambda,\hat\beta)$ are the power and thrust coefficients, respectively. 
At fixed blade pitch, the rotor speed affects both coefficients through the tip-speed ratio. The power coefficient determines the aerodynamic
power captured by the turbine and hence the aerodynamic torque, while the thrust coefficient determines the force exerted on the flow and the
resulting wake-velocity deficit.

We consider two aligned turbines separated by a fixed distance $L>0$. Turbine~1 is upstream and Turbine~2 is downstream. Both turbines have rotor radius $R$, diameter $D=2R$, swept area $A$, and fixed blade-pitch angle $\hat\beta$. The inertia, rotor speed, aerodynamic torque, and generator torque of Turbine~$i$ are denoted by $I_i$, $\omega_i(t)$, $T_{a,i}(t)$, and $T_{g,i}(t)$, respectively, for $i\in\{1,2\}$. Turbine~1 receives the constant free-stream inflow $v_1>0$, while Turbine~2 receives the wake-affected inflow $v_2(t)>0$.

Applying \eqref{Eqn:SingleTurbineDynamics} and \eqref{Eqn:SingleTurbineAerodynamicTorque} to the two turbines gives
\begin{subequations}\label{Eqn:TwoTurbineDynamics}
\begin{align}
I_1\dot{\omega}_1(t)&=\frac{\rho A v_1^3}{2\omega_1(t)}
C_P\!\left(\frac{R\omega_1(t)}{v_1},\hat\beta\right)-T_{g,1}(t),
\label{Eqn:UpstreamTurbineDynamics}\\
I_2\dot{\omega}_2(t)&=\frac{\rho A v_2^3(t)}{2\omega_2(t)}
C_P\!\left(\frac{R\omega_2(t)}{v_2(t)},\hat\beta\right)-T_{g,2}(t).
\label{Eqn:DownstreamTurbineDynamics}
\end{align}
\end{subequations}

The downstream inflow $v_2(t)$ couples the two rotor dynamics aerodynamically. A steady-state wake model computes $v_2(t)$ from the current upstream rotor speed $\omega_1(t)$, so it treats the effect of Turbine~1's operation on Turbine~2 as instantaneous. The PDE and
state-dependent delay models account for finite propagation: an effect of Turbine~1's operation at time $t_0$ reaches Turbine~2 at a later time
$t$. The interval $t-t_0$ depends on the wake velocity between the turbines. Since that velocity depends on the upstream thrust coefficient,
which depends on $\omega_1(t_0)$, the propagation delay depends on the upstream rotor speed.

The operating points $(\omega_1^*,\omega_2^*)$ specify the rotor speeds associated with steady-state power capture. Regulation about those points
depends on the generator-torque laws and the coupled rotor dynamics. Next, we consider generator-torque feedback based on rotor-speed deviations
to regulate turbine operation about the operating points.

\subsection{Regulating Turbines About Their Operating Points}

To regulate the turbines about their operating points $(\omega_1^*,\omega_2^*)$, we combine steady generator torques with feedback based on rotor-speed deviations. The steady generator torques balance the aerodynamic torques at the operating points:
\vspace{-2mm}
\begin{align}
T_{g,1}^*
&=\frac{P_a(\omega_1^*;v_1)}{\omega_1^*},
\label{Eqn:UpstreamEquilibriumTorque} \\
T_{g,2}^*
&=\frac{P_a\bigl(\omega_2^*;v_2(\omega_1^*)\bigr)}{\omega_2^*}.
\label{Eqn:DownstreamEquilibriumTorque}
\end{align}
 Applying local feedback to rotor speeds allows each generator torque to regulate to its own rotor-speed deviation. Since upstream operation affects downstream inflow, we also allow the upstream generator torque to regulate the speed deviation at the downstream turbine. We consider
\begin{equation}\label{Eqn:UpstreamTorqueFeedback}
T_{g,1}(t)=T_{g,1}^*+K_1\bigl(\omega_1(t)-\omega_1^*\bigr)
+K_2\bigl(\omega_2(t)-\omega_2^*\bigr),
\end{equation}
\begin{equation}\label{Eqn:DownstreamTorqueFeedback}
T_{g,2}(t)=T_{g,2}^*+K_3\bigl(\omega_2(t)-\omega_2^*\bigr),
\end{equation}
where $K_1,K_2,K_3\in\mathbb{R}$ are feedback gains. The gains $K_1$ and $K_3$ set the local torque response for Turbines~1 and~2, respectively. For positive $K_1$ and $K_3$, each local feedback term increases the generator torque when the corresponding rotor speed exceeds its operating speed, reducing rotor acceleration for a fixed aerodynamic torque. The gain $K_2$ coordinates upstream operation with the downstream rotor speed deviation.


Next, we combine the rotor-dynamics~\eqref{Eqn:TwoTurbineDynamics}, with the inflow velocity, $v_2$ as per the explicit state-dependent wake model~\eqref{Eqn:ExplicitSDModel} to obtain a state-dependent delay differential equation.

\section{Combination of Wake Model with Rotor Dynamics}\label{Sec:CoupledSystem}
The wake velocity at $x=0$ is determined by the current rotor speed of an upstream turbine, Turbine~1.
The actuator-disk theory gives the wake velocity immediately downstream of Turbine~1 as
\begin{equation}\label{Eqn:WakeVelocityBoundary}
\begin{aligned}
v_2(t,0)
&=v_1\left|1-2a_1\bigl(\omega_1(t)\bigr)\right| \\
&=v_1\sqrt{1-C_T\left(R\omega_1(t) /v_1 ,\hat\beta\right)}.
\end{aligned}
\end{equation}
Thus, a change in the upstream rotor speed changes the velocity of the wake entering the interval $[0,L]$.

\textbf{Model 5:}
Assuming fixed turbine spacing $L$. Substituting the inflow velocity $v_2$ from the explicit state-dependent delay wake model~\eqref{Eqn:ExplicitSDModel}  into rotor dynamics~\eqref{Eqn:TwoTurbineDynamics}, gives the SD-DDE:
\begin{equation}\label{Eqn:ProblemTwoTurbineModel}
\begin{aligned}
&I_1\dot{\omega}_1(t)=\frac{\rho A v_1^3}{2\omega_1(t)}
C_P\!\left(\frac{R\omega_1(t)}{v_1},\hat\beta\right)-T_{g,1}(t),\\
&I_2\dot{\omega}_2(t) =\frac{\rho A h^3(t)}{2\omega_2(t)}
C_P\!\left(\frac{R\omega_2(t)}{h(t)},\hat\beta\right)-T_{g,2}(t),\\
&h(t)=\beta(L)g\bigl(t-\tau(g(t))\bigr)+[1-\beta(L)]v_1
\end{aligned}
\end{equation}
with
\vspace{-2mm}
\[
g(t) = v_1\sqrt{1-C_T\left(\frac{R\omega_1(t)}{v_1},\hat\beta\right)},
\]
and generator-torque laws:
\begin{align*}
T_{g,1}(t)&=T_{g,1}^*+K_1\bigl(\omega_1(t)-\omega_1^*\bigr)
+K_2\bigl(\omega_2(t)-\omega_2^*\bigr), \\
T_{g,2}(t)&=T_{g,2}^*+K_3\bigl(\omega_2(t)-\omega_2^*\bigr),
\end{align*}
 are the coordinated generator-torque laws.

\section{Numerical Comparison of the Wake Models}\label{Sec:NumericalResults}

We compare the coupled rotor dynamics under four wake models: the steady-state Jensen--Park model~\eqref{Eqn:JensenWakeVelocity}, the PDE wake model~\eqref{Eqn:DynamicWakeVelocityPDE}, the forward-time Abel-type functional wake model~\eqref{Eqn:ForwardWakePropagation}, and the explicit state-dependent delay approximation of the wake~\eqref{Eqn:ExplicitSDModel}. Jensen--Park wake model evaluates inflow velocity at a downstream turbine using the current upstream rotor speed. The other three, dynamic models, models account for the time required for the wake to propagate between the turbines. We use the PDE response as the reference for comparing the transient behavior of the wake models coupled with the rotor dynamics~\eqref{Eqn:TwoTurbineDynamics}.

The simulations use the following parameters: $v_1=8\ \mathrm{m/s}$, $D=100\ \mathrm{m}$, $R=D/2$, $L=900\ \mathrm{m}$, $k_w=0.065$, $\rho=1.17\ \mathrm{kg/m^3}$, and $I_1=I_2=4\times10^7\ \mathrm{kg\,m^2}$. And, at the fixed blade-pitch angle, we consider aerodynamic functions:
\[
\begin{aligned}
C_P(\lambda,\hat\beta)&=0.4\exp\!\left[-\left(\frac{\lambda}{7}-1\right)^2\right],\\
a(\lambda)&=\frac{1}{3}+0.1\tanh\!\left(\frac{\lambda-7}{2}\right),\\
C_T(\lambda,\hat\beta)&=4a(\lambda)[1-a(\lambda)].
\end{aligned}
\]
The reference upstream speed is $\omega_1^*=7v_1/R$. The reference downstream inflow is obtained from the steady wake associated with $\omega_1^*$, and $\omega_2^*=7v_2^*/R$. We use the generator-torque laws~\eqref{Eqn:UpstreamTorqueFeedback} and~\eqref{Eqn:DownstreamTorqueFeedback} with $K_1=K_3=2\times10^5$ and $K_2=-10^7$. 
The rotor-speed references and feedback gains are identical across the four models.

Define
\[
x_i(t):=\omega_i(t)-\omega_i^*,
\qquad i\in\{1,2\},
\]
where $x_i$ is the rotor-speed deviation of Turbine~$i$. We consider $x_1(0)=x_2(0)=10^{-2}\ \mathrm{rad/s}$ and $x_1(0)=x_2(0)=1\ \mathrm{rad/s}$. 
For each case, rotor histories are held at their initial speeds for $t\leq0$. The PDE starts from the steady wake associated with the initial upstream speed:
\[
v_2(0,x)=\beta(x)v_2(0,0)+[1-\beta(x)]v_1,
\qquad x\in[0,L],
\]
where $\beta$ is defined by \eqref{Defn:Beta} and the boundary velocity $v_2(0,0)$ is given by \eqref{Eqn:WakeVelocityBoundary}. 

Since the turbine spacing is fixed, write $v_2(t):=v_2(t,L)$. 
For the forward-time Abel-type functional wake model, we consider the seed function
\begin{equation}\label{Eqn:FunctionalWakeInitialSegment}
\phi(t)=\beta(L)v_2(0,0)+[1-\beta(L)]v_1,
\qquad 0\leq t\leq\tau(v_2(0,0)),
\end{equation}
which is recursively propagated forward in time to obtain the obtain the solution to the Abel-type functional equation~\cite{11615921}.
And for the explicit state-dependent delay model, Newton iteration solves at subsequent times:
\begin{equation}\label{Eqn:NumericalWakeInverse}
t_0+\tau(v_2(t_0,0))=t.
\end{equation}
Then upstream rotor speeds at $t_0$ are interpolated from  their stored history and inserted into the boundary condition~\eqref{Eqn:WakeVelocityBoundary}. Lem.~\ref{Lem.3} then gives
\vspace{-2mm}
\begin{equation}\label{Eqn:NumericalFunctionalWake}
v_2(t)=\beta(L)v_2(t_0,0)+[1-\beta(L)]v_1.
\end{equation}

The exact relation obtained by inverting $\theta$:
\begin{equation}\label{Eqn:NumericalImplicitWakeDelay}
\theta^{-1}(t)=t-\tau\bigl(v_2(\theta^{-1}(t),0)\bigr).
\end{equation}

The FIFO condition makes $\theta$ invertible. And  the slow-variation assumption supports replacing $\tau(v_2(\theta^{-1}(t),0))$ by $\tau(v_2(t,0))$ in \eqref{Eqn:NumericalImplicitWakeDelay}.

\begin{figure}[!t]
\centering
{\setlength{\tabcolsep}{3pt}\begin{tabular}{@{}ll@{}}
\tikz[baseline=-0.5ex]{\draw[blue,dashed,line width=0.9pt] (0,0)--(0.5,0);}
Jensen-Park model  &
     \tikz[baseline=-0.5ex]{\draw[black,line width=0.9pt] (0,0)--(0.5,0);}
PDE model \\
\tikz[baseline=-0.5ex]{\path[use as bounding box] (0,-0.06) rectangle (0.5,0.06); \draw[red,fill=white,line width=0.8pt] (0.25,0) circle (1.5pt);}
Abel-type model &
\tikz[baseline=-0.5ex]{\draw[wakegreen,densely dotted,line width=0.9pt] (0,0)--(0.5,0);}
Explicit state-dependent 
\end{tabular} \par}
    \begin{subfigure}[t]{0.49\linewidth}
        \centering
        \includegraphics[width=\linewidth]{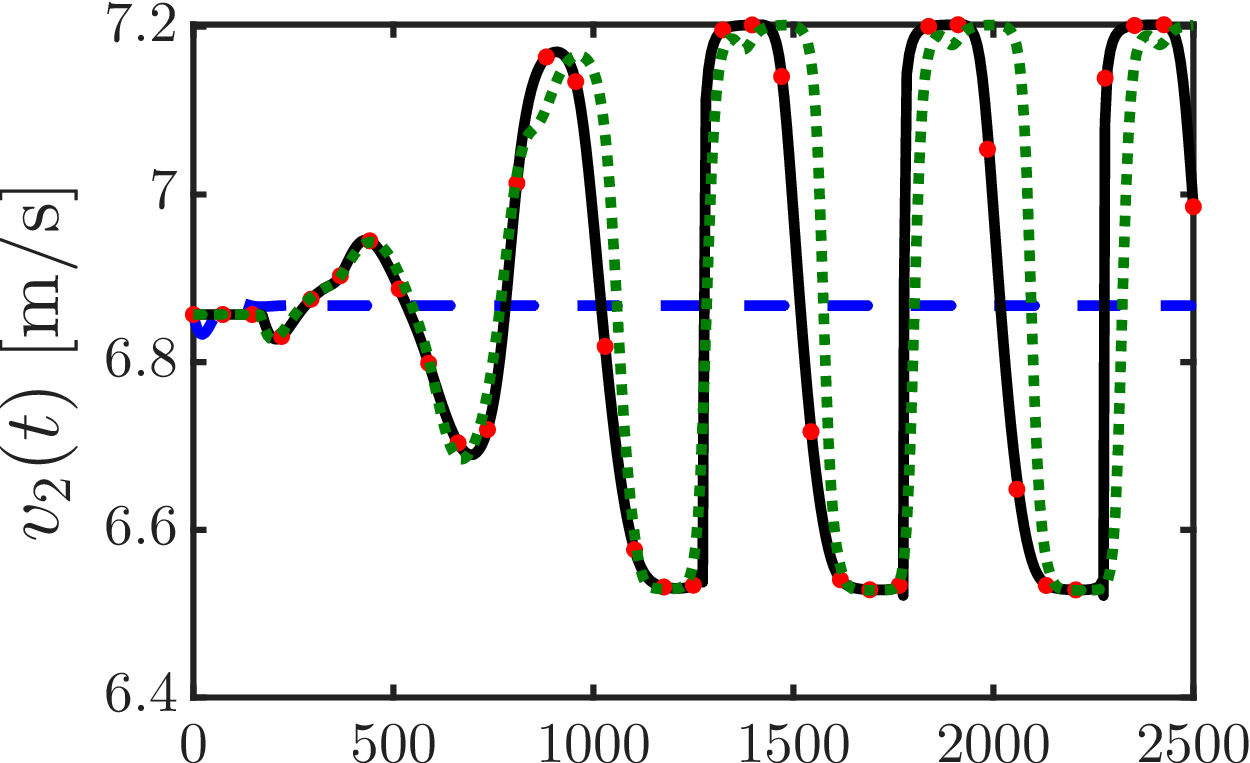}
        \subcaption{}
        \label{fig:two-up:g}
    \end{subfigure}\hfill
    \begin{subfigure}[t]{0.49\linewidth}
        \centering
        \includegraphics[width=\linewidth]{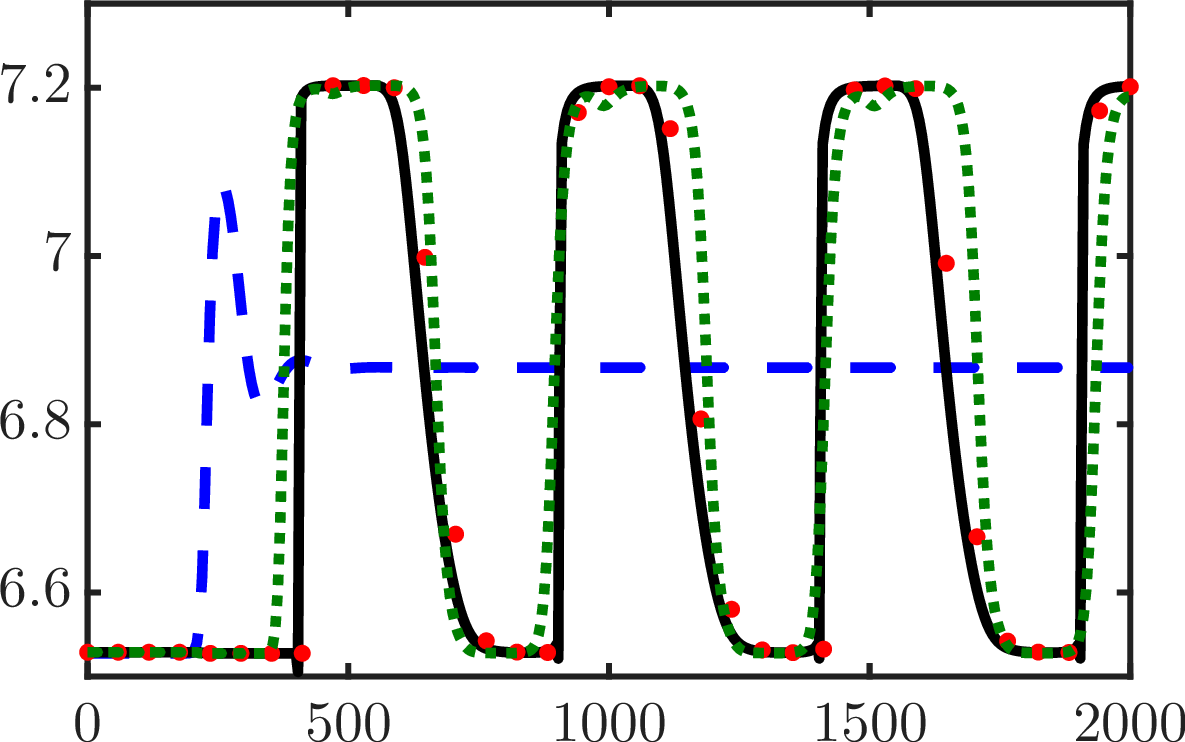}
        \subcaption{}
        \label{fig:two-up:h}
    \end{subfigure}

    \begin{subfigure}[t]{0.49\linewidth}
        \centering
        \includegraphics[width=\linewidth]{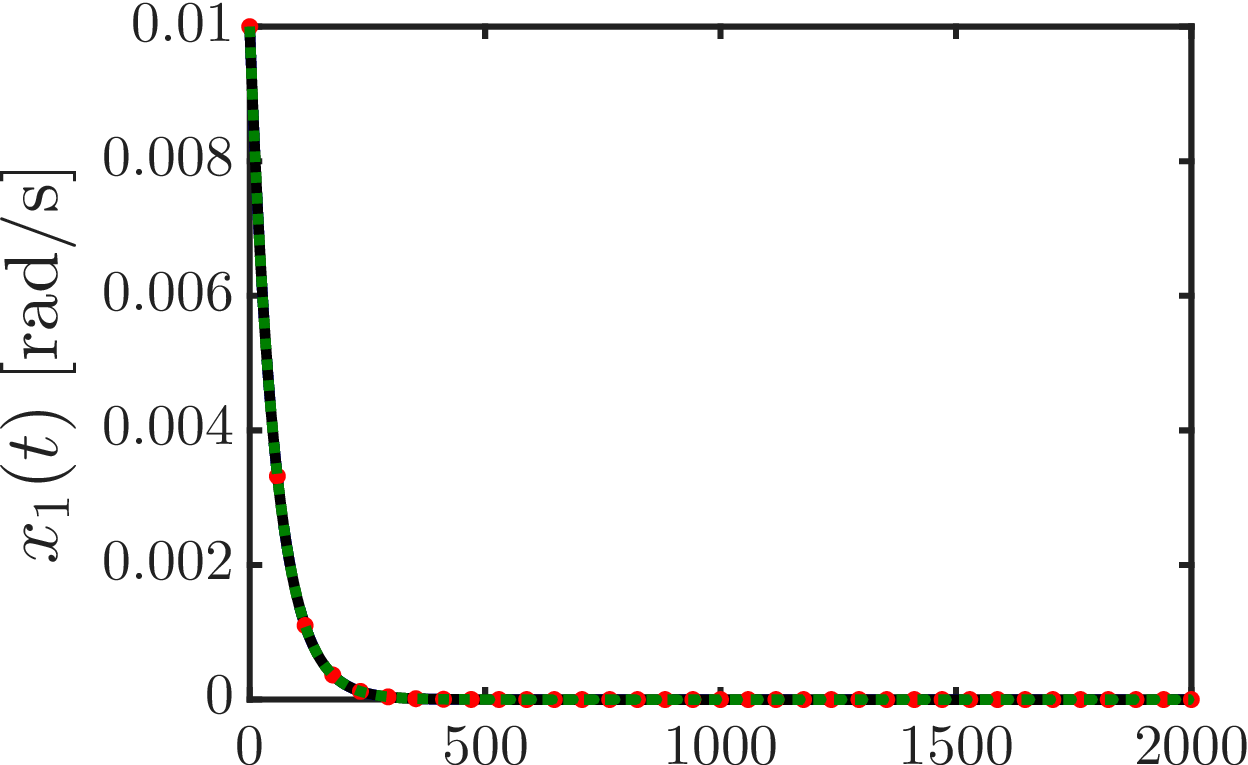}
        \subcaption{}
        \label{fig:two-up:i}
    \end{subfigure}\hfill
    \begin{subfigure}[t]{0.49\linewidth}
        \centering
        \includegraphics[width=\linewidth]{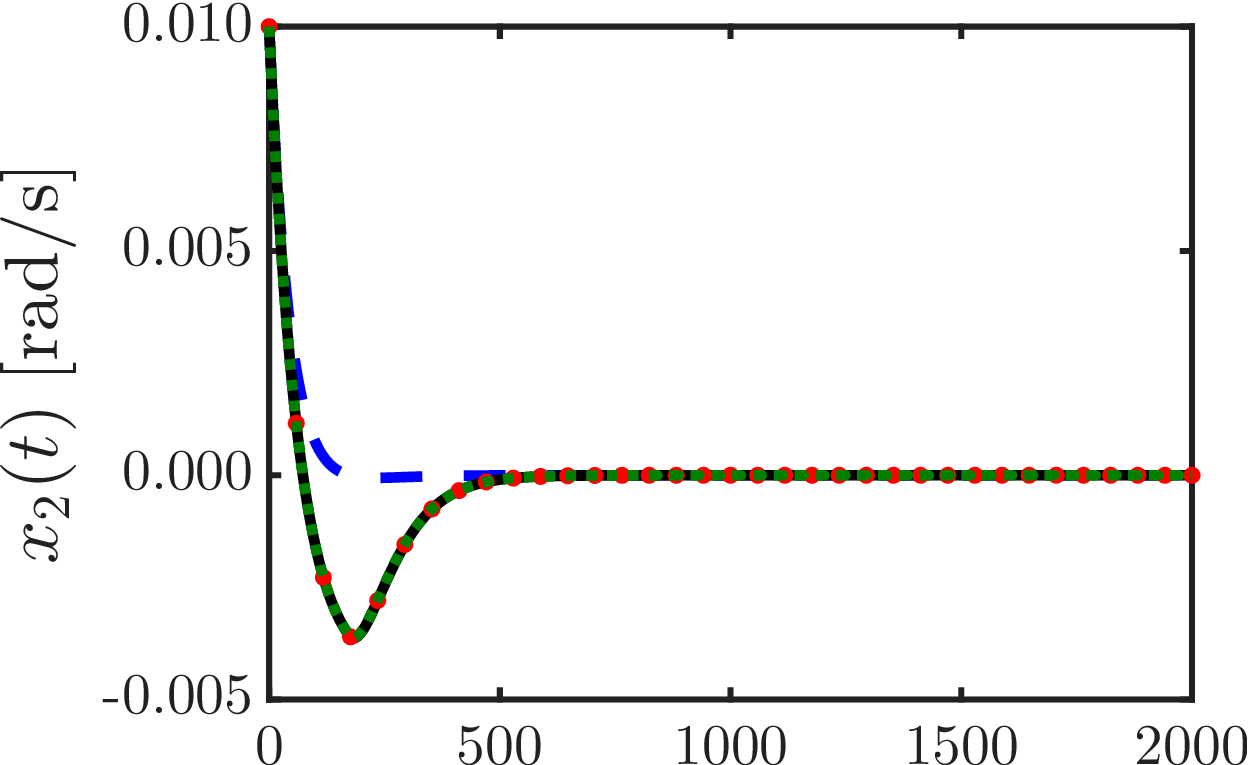}
        \subcaption{}
        \label{fig:two-up:j}
    \end{subfigure}
    
    \begin{subfigure}[t]{0.49\linewidth}
        \centering
        \includegraphics[width=\linewidth]{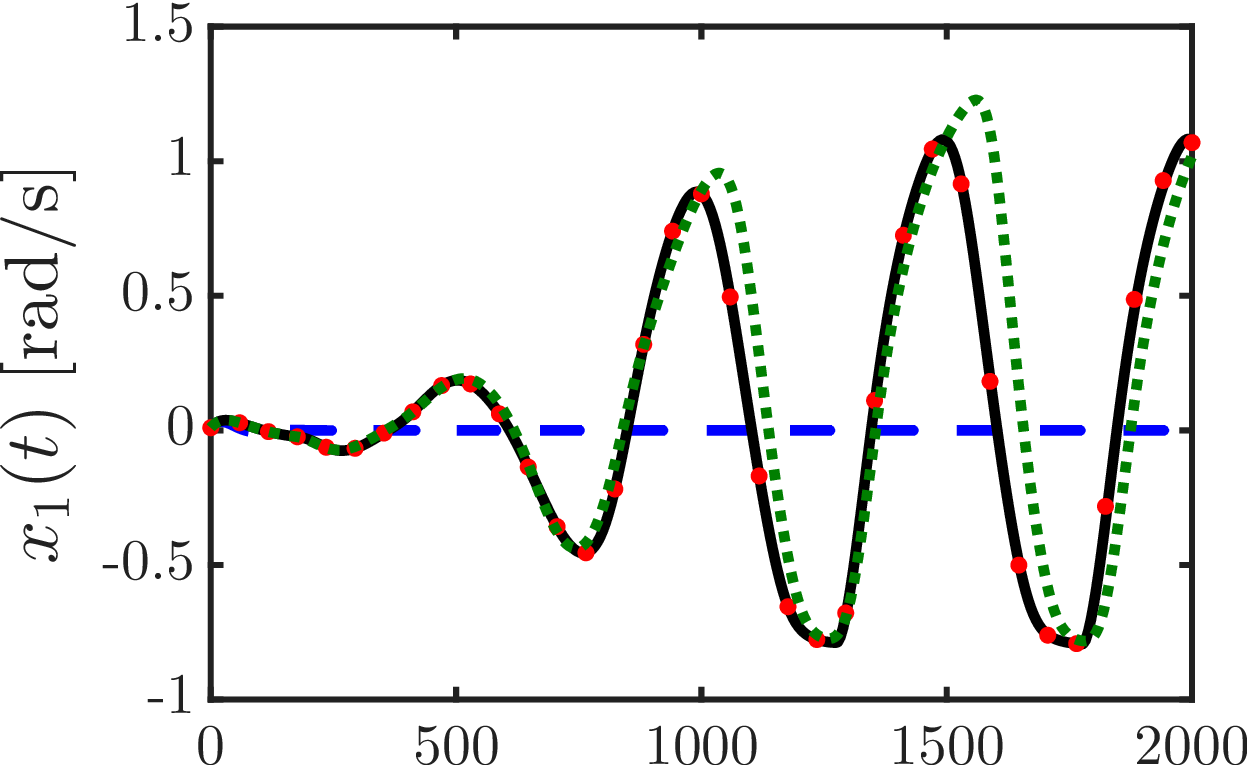}
        \subcaption{}
        \label{fig:two-up:c}
    \end{subfigure}\hfill
    \begin{subfigure}[t]{0.49\linewidth}
        \centering
        \includegraphics[width=\linewidth]{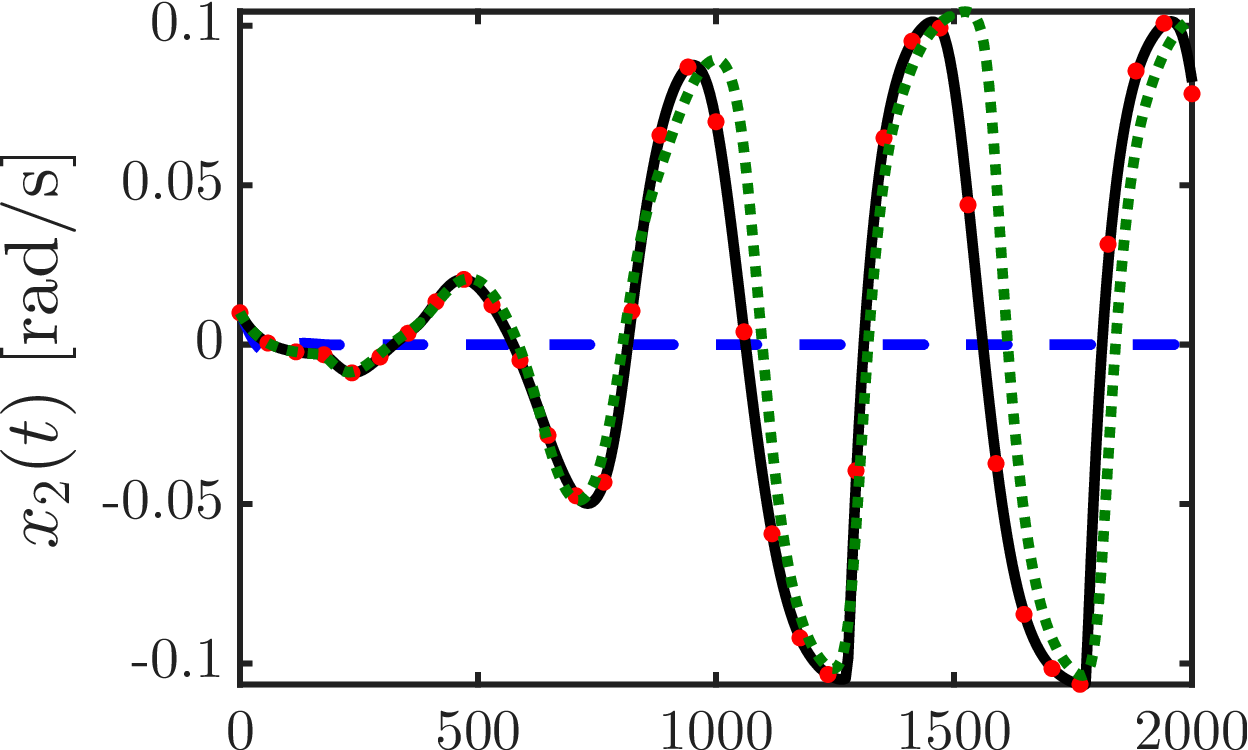}
        \subcaption{}
        \label{fig:two-up:d}
    \end{subfigure}

    \begin{subfigure}[t]{0.49\linewidth}
        \centering
        \includegraphics[width=\linewidth]{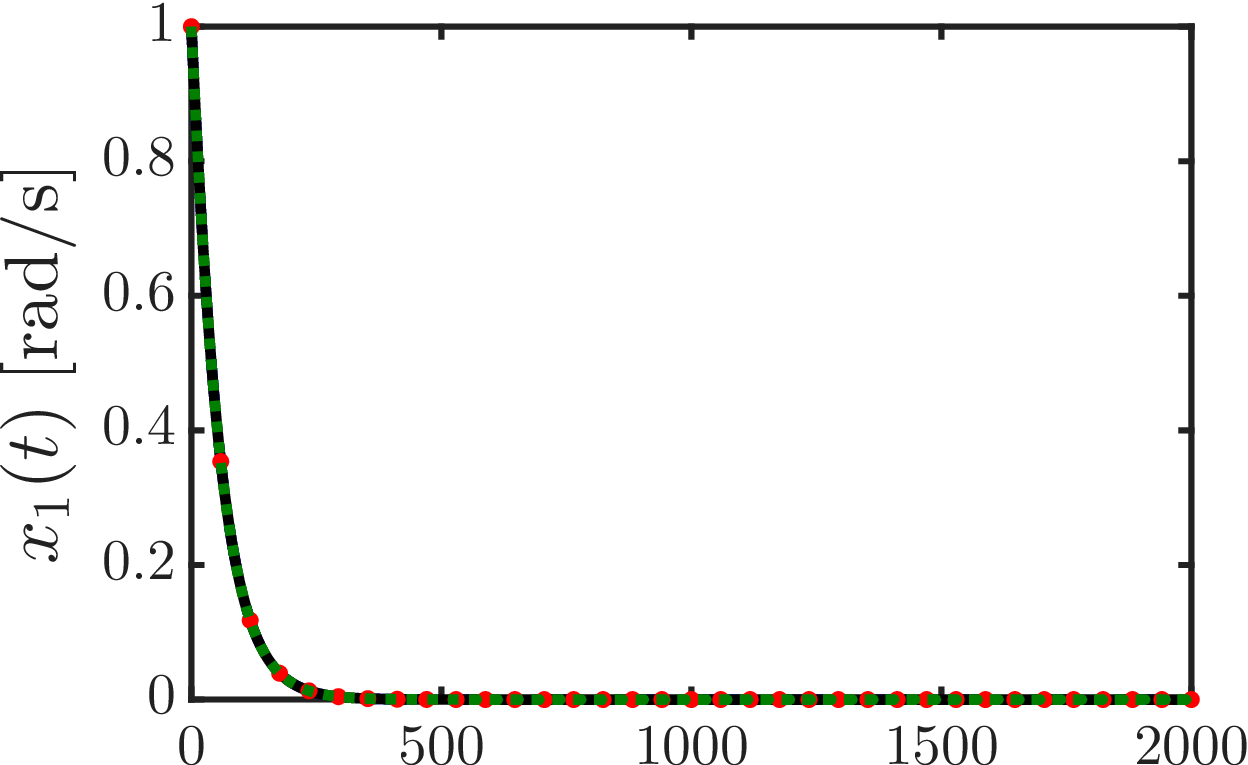}
        \subcaption{}
        \label{fig:two-up:a}
    \end{subfigure}\hfill
    \begin{subfigure}[t]{0.49\linewidth}
        \centering
        \includegraphics[width=\linewidth]{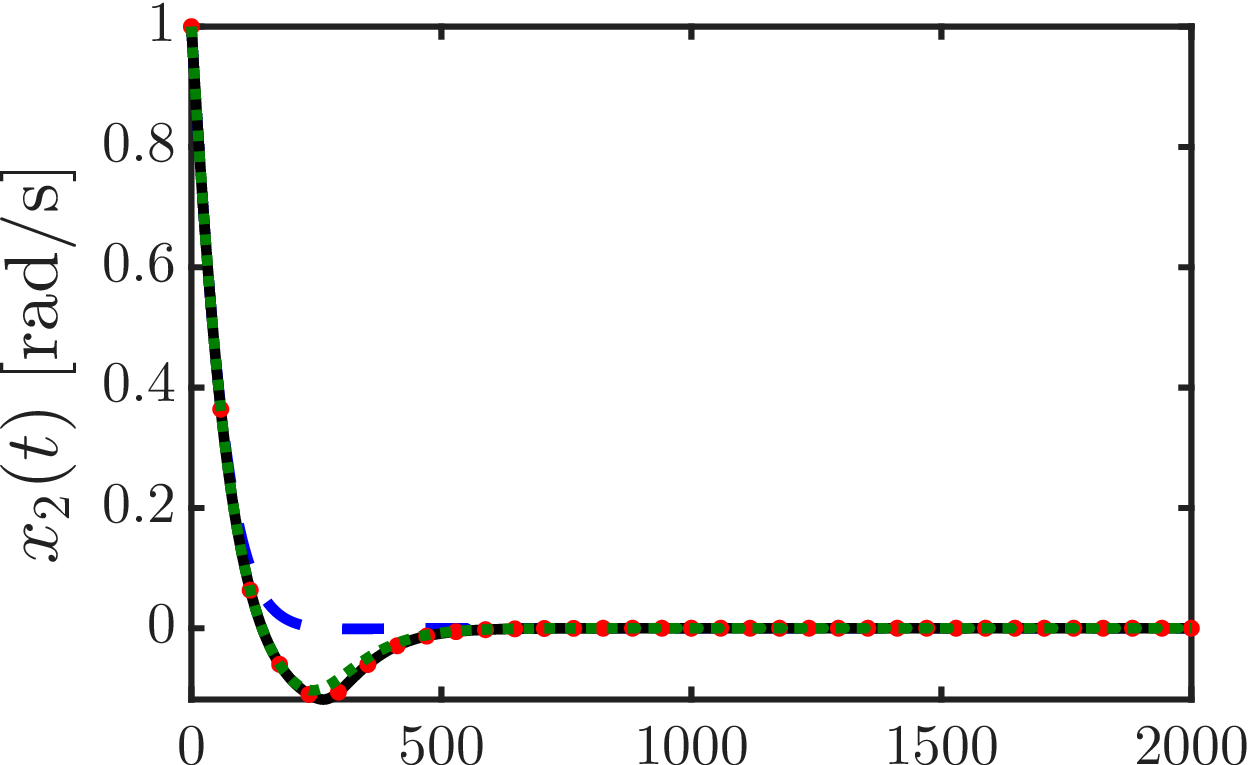}
        \subcaption{}
        \label{fig:two-up:b}
    \end{subfigure}

    \begin{subfigure}[t]{0.49\linewidth}
        \centering
        \includegraphics[width=\linewidth]{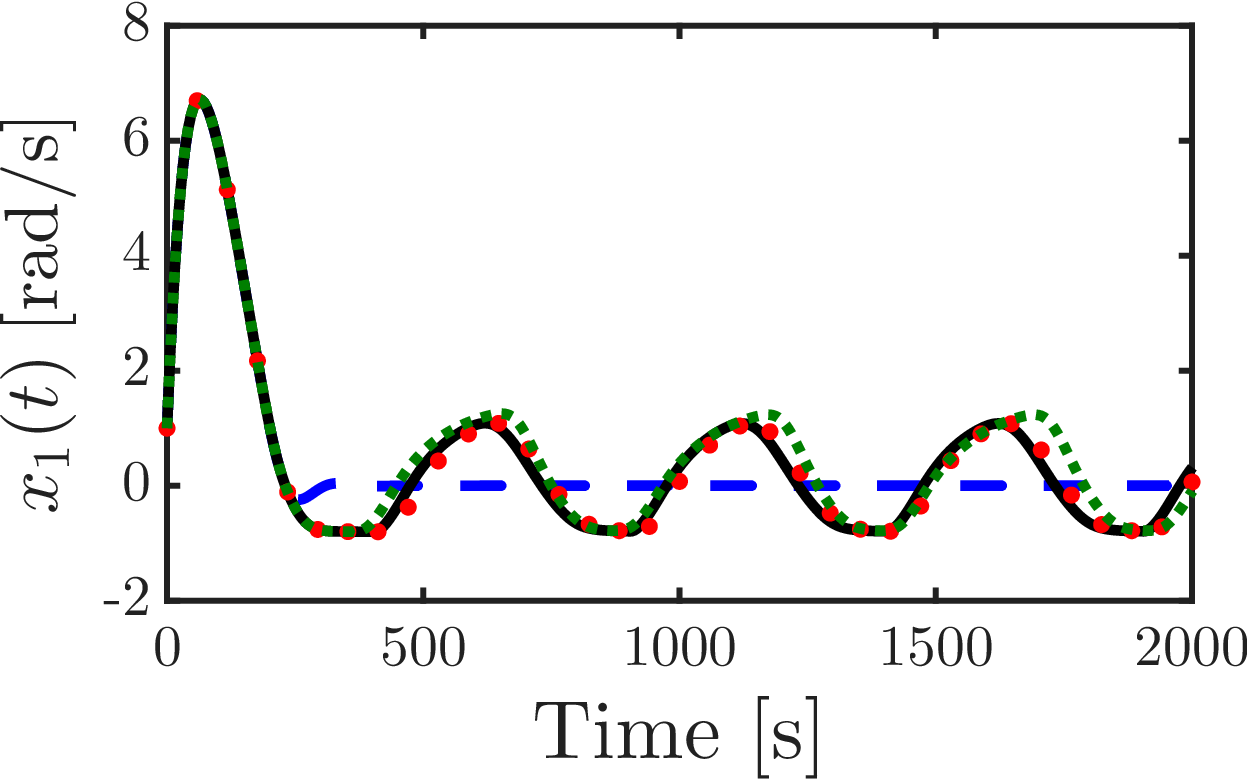}
        \subcaption{}
        \label{fig:two-up:e}
    \end{subfigure}\hfill
    \begin{subfigure}[t]{0.49\linewidth}
        \centering
        \includegraphics[width=\linewidth]{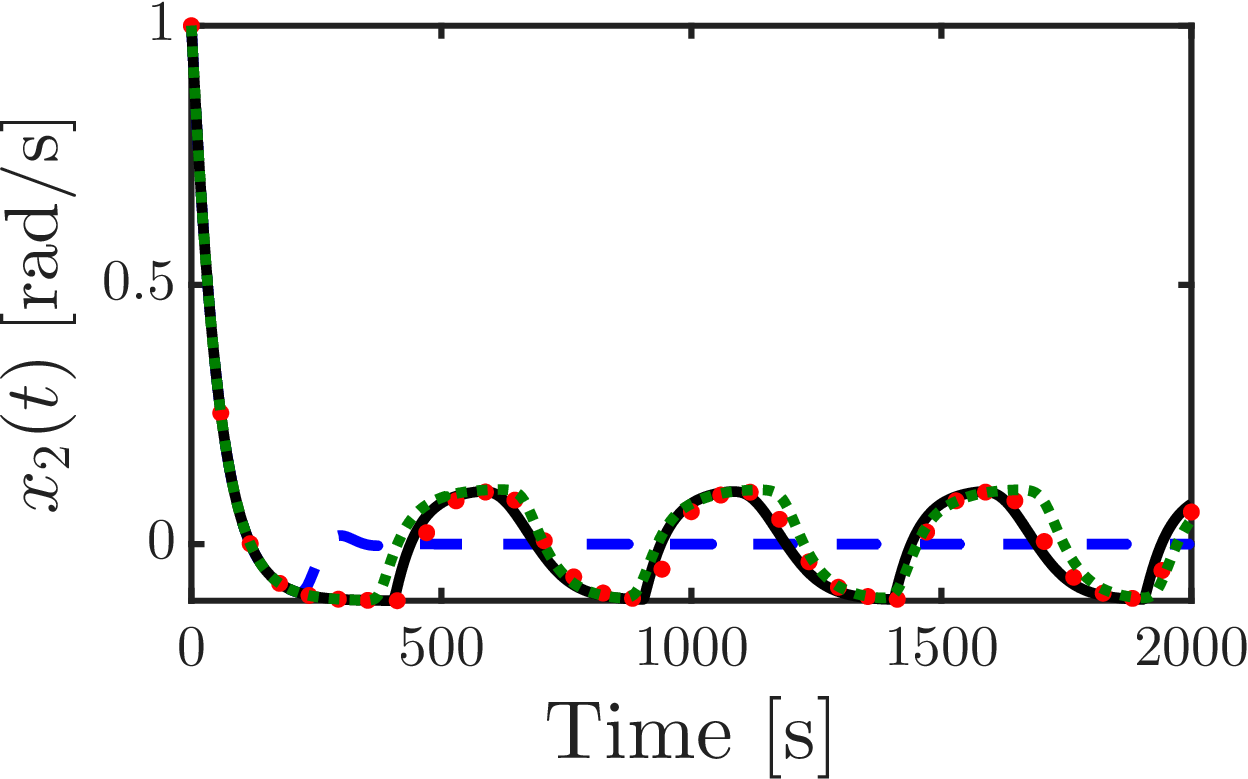}
        \subcaption{}
        \label{fig:two-up:f}
    \end{subfigure}

\caption{\small Open- and closed-loop comparison of  wake models. The closed-loop downstream inflow $v_2(t)$ for $x_1(0)=x_2(0)=10^{-2}\,\mathrm{rad/s}$ and $x_1(0)=x_2(0)=1\,\mathrm{rad/s}$ is shown in (a) and (b), respectively. The open-loop responses $x_1(t)$ and $x_2(t)$ are shown in (c) and  (d), respectively, and converge to the open-loop equilibrium for all four models. The closed-loop rotor-speed deviations are shown in (e) and (f) for $x_1(0)=x_2(0)=10^{-2}\,\mathrm{rad/s}$. For $x_1(0)=x_2(0)=1\,\mathrm{rad/s}$, the open loop responses are shown  in (g) and (h) and the closed loop responses in (i) and (j). The steady-state Jensen--Park response converges to equilibrium, whereas the PDE and forward-time Abel-type functional equation model responses exhibit repeated oscillations. The forward-time Abel-type functional equation model closely reproduces the PDE response, including the rapid transitions in $v_2(t)$. The explicit state-dependent delay approximation captures the oscillations but develops some drift in their timing.}
\label{fig:example1}
\end{figure}

\subsection{State-Dependent Delay Model Captures Wake-Transport Effects}

With feedback gains $K_1=K_3=2\times10^5$ and $K_2=-10^7$, the PDE and forward-time Abel-type functional wake models produce downstream inflow responses $v_2(t)$ that exhibit repeated oscillations after the initial transient, as shown in Fig.~\ref{fig:example1}(\subref{fig:two-up:g}) for the initial conditions $x_1(0)=x_2(0)=10^{-2}\ \mathrm{rad/s}$ and Fig.~\ref{fig:example1}(\subref{fig:two-up:h}) for the initial conditions $x_1(0)=x_2(0)=1\ \mathrm{rad/s}$. The steady-state Jensen--Park response does not exhibit these repeated oscillations because it does not account for the wake-transport delay. The Abel-type functional wake model closely follows the PDE response, including the rapid transitions in downstream inflow shown in Fig.~\ref{fig:example1}(b). The explicit state-dependent-delay approximation retains the repeated oscillations in $v_2(t)$ but differs from the PDE model near those transitions. These responses show that the delay models capture the wake-transport effects in $v_2(t)$, including oscillatory behavior absent from the steady-state Jensen--Park model, while the functional wake model more closely reproduces the PDE response.
\subsection{Destabilizing Effect of Wake-Transport Delay}
Figures~\ref{fig:example1}(\subref{fig:two-up:i}), \ref{fig:example1}(\subref{fig:two-up:j})  and Fig.~\ref{fig:example1}(\subref{fig:two-up:a}), \ref{fig:example1}(\subref{fig:two-up:b}) show the open-loop rotor-speed responses. For all four wake models, the responses converge to the open-loop equilibrium without sustained oscillations, showing that the open-loop systems are stable. With feedback gains $K_1=K_3=2\times10^5$ and $K_2=-10^7$, the responses in Fig.~\ref{fig:example1}(\subref{fig:two-up:c})--(\subref{fig:two-up:f}) show that wake-transport delay can destabilize dynamics that remain stable under the Jensen--Park model. For $x_1(0)=x_2(0)=10^{-2}\ \mathrm{rad/s}$, the Jensen--Park deviations in Fig.~\ref{fig:example1}\subref{fig:two-up:c} and Fig.~\ref{fig:example1}(\subref{fig:two-up:d}) decay toward zero, while the PDE and forward-time Abel-type functional equation model responses develop oscillations of increasing amplitude. The responses in Fig.~\ref{fig:example1}(\subref{fig:two-up:e}) and Fig.~\ref{fig:example1}(\subref{fig:two-up:f}) show the same contrast following a larger initial condition, $x_1(0)=x_2(0)=1 \; \mathrm{rad/s}$. For the Jensen--Park, rotor speed deviations from the operating points decrease toward zero, whereas the PDE and forward-time Abel-type functional equation model responses retain repeated oscillations. The responses of the forward-time Abel-type functional equation model closely follow the PDE responses for both turbines. The explicit state-dependent delay approximation captures the oscillations, although a difference in their timing becomes more apparent later in the simulation.

The stable open-loop responses show that the closed-loop oscillations result from the interaction between feedback and delayed aerodynamic coupling rather than from unstable open-loop rotor dynamics. The Jensen--Park model uses the current upstream rotor speed to determine the downstream inflow and therefore omits the wake-transport delay. In the PDE model, the downstream inflow depends on the upstream operation at an earlier time. Under the same torque laws and feedback gains, this delay shifts the timing of the interaction between the turbines and causes the rotor-speed oscillations to grow rather than decay. The close agreement between the PDE and forward-time Abel-type functional equation model responses supports this behavior to wake propagation, while the explicit state-dependent delay approximation reproduces the oscillations with some drift in their timing. Thus, although the open-loop system is stable, the transport delay introduced by wake propagation destabilizes the closed-loop dynamics and drives the rotor-speed deviations away from the operating point.

\section{Conclusion}

Wake transport can destabilize the operating point of two coupled turbines because changes in upstream operation reach the downstream turbine after a state-dependent propagation time. For the wake PDE~\eqref{Eqn:ProblemPDE}, integration along characteristic curves gives an Abel-type functional equation~\eqref{Eqn:ProblemForwardPropagation}. We then assume first-in first-out (FIFO) ordering to obtain an implicit state-dependent delay wake model~\eqref{Eqn:ProblemStateDependentWake}. Furthermore, we assume that the transport delay due wake interactions is slow-varying, to obtain an approximate explicit state-dependent delay model of the wake propagation~\eqref{Eqn:ExplicitWakeProb}.

Numerical comparison of the wake models shows that the steady-state wake model converges under feedback gains for which the PDE wake model exhibits destabilizing oscillations. The Abel-type functional model closely follows the PDE response, while the explicit state-dependent delay model captures the oscillations with some differences in phase and amplitude. The explicit state-dependent thus retain the effect of wake propagation needed  for stability analysis and control design.

\bibliographystyle{IEEEtran}
\bibliography{References}
\end{document}